\documentclass[aps,floats,epsf, twocolumn ]{revtex4}
 \usepackage{graphics}

\begin{document}

\title{Unconventional superconductivity on honeycomb lattice: the theory of  Kekule order parameter}
\author{Bitan Roy and Igor F. Herbut}

\affiliation{Department of Physics, Simon Fraser University,
 Burnaby, British Columbia, Canada V5A 1S6}

\begin{abstract}
 A  spatially non-uniform superconducting phase is proposed as the electronic variational ground state for the attractive interactions between nearest neighbors on graphene's honeycomb lattice, close to and right at the filling one half. The state spontaneously breaks the translational invariance of the lattice into the Kekule pattern of bond order parameters, and it is gapped, spin triplet, and odd under the sublattice exchange. With the increase of attractive interactions we first find the transition from the semimetallic phase into the p-Kekule superconductor, defined as being odd under the exchange of Dirac points, with the additional discontinuous superconductor-superconductor transition into the even s-Kekule state, deep within the superconducting phase. Topological excitations of the Kekule superconductor and its competition with other superconducting states on the honeycomb lattice are discussed.

\end{abstract}
\maketitle

\vspace{10pt}

\section{Introduction}

Fermions on graphene's honeycomb lattice can in principle find themselves in a plethora of insulating phases, depending on the relative magnitudes of different components of a finite-range repulsive interaction, for example \cite{herb-jur-roy}. If the net interaction would have an attractive component, on the other hand, there would be a variety of superconducting states available to Dirac quasiparticles for pairing and condensation. Some of them are quite conventional: the on-site attraction would clearly favor the usual s-wave singlet pairing \cite{zhao}. Others are already less so; the second-nearest-neighbor attraction, for example, leads to an f-wave superconductor \cite{honerkamp}, which changes sign six times around the Brillouin zone. Another exotic superconducting state on honeycomb lattice was argued to arise from  the nearest-neighbor attraction \cite{uchoa}: instead of gapping the Dirac points it lowers the energy of the Dirac-Fermi sea by effectively increasing the Fermi velocity. Only away from  half-filling does this state acquire a finite superconducting gap, which is then proportional to the chemical potential. A closely related superconducting ground state was also discussed in the context of  the t-J-U model and graphite \cite{annica}. This {\it hidden}  superconducting order is otherwise a spin-singlet, and even under the exchange of the two sublattices and/or the Dirac points. Since the electrons in graphene have three sets of discrete indices, the sublattice, valley, and real spin, possible superconducting states may exhibit various symmetries with respect to spatial and time inversions \cite{dima, herbut}. Together with the observation of superconductivity in graphite \cite{kopelevich}, the intricate structure of the superconducting vortex \cite{wilczek, herbut, ghaemi}, novel proximity phenomena \cite{benakker} and the quantum  criticality  \cite{herb-jur-vaf, metzner}, this makes the problem of superconductivity in graphene or in an optical honeycomb lattice engaging from theoretical as well as experimental  points of view.

In this paper we will be concerned with the forms of the superconducting condensate on the honeycomb lattice at, and therefore also near, half-filling. As a convenient point of departure we consider the problem of graphene with the chemical potential right at the Dirac point and with the attraction only between the electrons residing on the nearest neighbors of the honeycomb lattice. The motivation for studying such a pairing interaction of a finite range comes in part from the theories of boson-fermion mixtures in optical lattices, where the nearest-neighbor attraction between fermions arises upon integration over the bosonic degrees of freedom \cite{smith}. Also, since the fermions in reality certainly experience a strong repulsion when they find themselves on the same site, the attraction between the nearest neighbors appears to be the simplest reasonable assumption that would still lead to pairing. Our conclusion about the superconducting ground state that arises as the BCS mean-field solution in this model is unusual and qualitatively different from the previous study \cite{uchoa}. Within the standard mean-field approach we find the superconducting state with the lowest energy to be the spin-triplet, {\it non-uniform} condensate, which is odd under the exchange of the two sublattices. The spatial Kekule pattern \cite{hou} of bonds between the paired electrons on nearest-neighbors has the periodicity of $2\vec{Q}$, where $\pm \vec{Q}$ are the Dirac points, which allows it to connect the two Dirac valleys and that way open the mass-gap in the Bogoliubov quasiparticle spectrum. It is an example of Fulde-Ferrell-Larkin-Ovchinikov  \cite{fulde, larkin} type of superconducting phase appropriate to the honeycomb lattice.

We argue that the  development of such a non-uniform superconductor at T=0 may preempt the formation of the previously proposed hidden order, which in our approximation we indeed find to be suppressed at all couplings. The Kekule superconductor breaks the exact particle-number and the spin-rotational symmetries, and exhibits three massless and three massive modes in the ordered phase. It also rather weakly breaks the internal and approximate $U(1)$ symmetry between various Kekule patterns. Near the semimetal-superconductor transition we find the p-Kekule state, odd under the valley exchange, to have the lowest energy, with an additional discontinuous transition {\it within} the superconducting phase into the s-Kekule state, even under the valley exchange, at a stronger attractive interaction.

 The target space of the Kekule order parameter is $S_3$, the surface of sphere in four dimensions. The topology of this space implies that there are no stable topological defects in our two-dimensional system, and therefore presumably no sharp finite temperature phase transition. Explicit breaking of the rotational symmetry, by an external magnetic field or the spin-orbit interaction, for instance, changes the target space for the order parameter and restores the possibility of topologically distinct defects. The cases of easy plane and easy axis, introduced by the two terms mentioned above, are both discussed. We also list all other gapped and hidden (gapless) superconducting states on the honeycomb lattice, and briefly comment on  their competition.

 The paper is organized as follows. In the next section we write the Bogoliubov-de Gennes Hamiltonian in the Dirac form, and introduce the non-uniform Kekule ansatz for the superconducting bond order parameters. The minimization of the mean-field energy for the simplest non-uniform state and the resulting s-Kekule superconductor is presented in section III. In sec. IV we discuss the competition between the Kekule and hidden orders. In sec. V a more general Kekule pattern is considered, and the p-Kekule state is defined. The other possible  superconducting orders are discussed in sec. VI, and the issue of topological defects and the target space for the Kekule order parameter in sec. VII. Concluding remarks are given in sec. VIII.

 \begin{figure}[t]
{\centering\resizebox*{60mm}{!}{\includegraphics{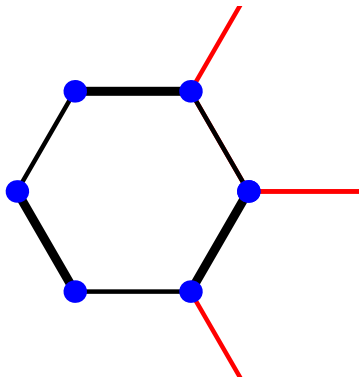}}
\par} \caption[] {The unit cell of the Kekule lattice of superconducting bond order parameters. The red line corresponds to $\Delta \cos \alpha$, the bold line to $\Delta \cos( \alpha + 2\pi/3)$, and the thin line to $\Delta \cos(\alpha - 2\pi/3 )$. The unit cell contains six sites (blue points) and nine bonds. When periodically arranged in a triangular lattice of period $3$ it yields the Kekule pattern.}
\end{figure}

\section{BdG-Dirac Hamiltonian and the Kekule ansatz}

  Consider the usual tight-binding Hamiltonian for spin-1/2 fermions on honeycomb lattice at half-filling, with an attractive interaction between the nearest neighbors,
\begin{equation}
H= H_t  - V \sum_{ \langle \vec{x},\vec{y} \rangle} \sum_{ \sigma, \sigma ' =\uparrow, \downarrow } n_\sigma (\vec{x}) n_{\sigma '} (\vec{y}),
\end{equation}
\begin{equation}
H_t = t \sum_{ \langle \vec{x},\vec{y} \rangle} \sum_{\sigma =\uparrow, \downarrow } u_\sigma ^\dagger (\vec{x}) v_\sigma (\vec{y}) +h.c.,
\end{equation}
where $V>0$. $u_\sigma (\vec{x})$ and $v_\sigma (\vec{y})$ are the fermionic operators at the two triangular sublattices of the honeycomb lattice. Decoupling the interaction term in the particle-particle channel yields the Bogoliubov-de Gennes (BdG) Hamiltonian
\begin{equation}
H_{BdG} = H_t - \sum_{ \langle \vec{x},\vec{y} \rangle} \sum_{ \sigma, \sigma ' =\uparrow, \downarrow } \Delta_{\sigma \sigma'} (\vec{y},\vec{x}) u_{\sigma'}^\dagger (\vec{x}) v_{\sigma } ^\dagger (\vec{y}) + h. c. ,
\end{equation}
with the superconducting order parameters to be determined self-consistently as
\begin{equation}
\Delta_{\sigma \sigma '} (\vec{y},\vec{x}) = V \langle v_\sigma (\vec{y}) u_{\sigma '} (\vec{x}) \rangle.
\end{equation}

   Assuming the order parameters to be  much smaller than the bandwidth the condensation energy comes mainly from the pairing of the quasiparticle states near the two Dirac points. Let us form a 16-component Dirac-Nambu fermion  $\Psi = (\Psi_p, \Psi_h)^\top$, with
   $\Psi_p = (\Psi_{p\uparrow}, \Psi_{p\downarrow})^\top $ and $\Psi_h = (\Psi_{h\downarrow} , - \Psi_{h\uparrow})^\top$, and
 \begin{equation}
  \Psi^\top _{p \sigma}  (\vec{q}) =
  ( u _\sigma (\vec{Q} + \vec{q}), v _\sigma (\vec{Q}+ \vec{q}), u_\sigma (-\vec{Q}+ \vec{q}), v_\sigma (-\vec{Q}+ \vec{q}) ),
  \end{equation}
  \begin{equation}
  \Psi^\top _{h \sigma}  (\vec{q}) =
  ( v_\sigma ^\dagger ( \vec{Q} - \vec{q}), u _\sigma ^\dagger (\vec{Q} - \vec{q}), v_\sigma ^\dagger (-\vec{Q}- \vec{q}), v_\sigma ^\dagger (-\vec{Q}- \vec{q}) ).
  \end{equation}
  The tight-binding Hamiltonian at low energies then becomes
  \begin{equation}
  H_t = \sum_{\vec{q}} \Psi^\dagger  (\vec{q}) H_D  \Psi(\vec{q}) +O(q^2),
  \end{equation}
  with $H_D$ as the Dirac Hamiltonian in two dimensions, which in our construction and in the first quantization assumes a particularly simple form,
  \begin{equation}
  H_D=\tau_0 \otimes \sigma_0 \otimes i \gamma_0 \gamma_i q_i.
  \end{equation}
   Here, $\gamma_0= \sigma_0 \otimes \sigma_3$, $\gamma_1= \sigma_3 \otimes \sigma_2$, $\gamma_2= \sigma_0 \otimes \sigma_1$, are the usual four-component anticommuting Hermitian gamma-matrices \cite{herb-jur-roy}. The two-component Pauli matrices $\{ \tau_0, \vec{\tau} \}$ operate on Nambu's, and $\{ \sigma_0, \vec{\sigma} \}$ on the spin indices.  We will also define the remaining two gamma-matrices as $\gamma_3= \sigma_1 \otimes \sigma_2$, and $\gamma_5= \sigma_2 \otimes \sigma_2$. For convenience, hereafter we also set the Fermi velocity $v_F=\sqrt{3}t/2=1$ and the lattice spacing $a$ to unity.

Next, we define the {\it Kekule ansatz} for the superconducting order parameter:
  \begin{equation}
  \Delta_{\sigma\sigma} (\vec{x},\vec{y} ) = \Delta_{ \sigma } \cos( \vec{Q} \cdot (\vec{x} + \vec{y})+\alpha ),
  \end{equation}
  \begin{equation}
  \frac{1}{2} (\Delta_{\downarrow\uparrow}(\vec{x},\vec{y} ) + \Delta_{\uparrow\downarrow} (\vec{x},\vec{y} ) )  = \Delta \cos( \vec{Q} \cdot (\vec{x} + \vec{y})+ \alpha ),
  \end{equation}
  \begin{equation}
  \frac{1}{2} ( \Delta_{\downarrow\uparrow}(\vec{x},\vec{y}) - \Delta_{\uparrow\downarrow} (\vec{x},\vec{y}) ) =  \Delta '.
  \end{equation}
  The components of the triplet are assumed to be spatially periodic, with the periodicity of $2\vec{Q}$, whereas the singlet component is simply uniform. The ``angle" $\alpha$ parameterizes different spatial patterns of the order parameter. The unit cell of the Kekule lattice is depicted in Fig. 1.

\section{s-Kekule ground state}

  We determine first the optimal Kekule ground state for $\alpha=0$, and then consider a more general solution.
  With the above ansatz the  BdG Hamiltonian can be rewritten as
 \begin{equation}
 H_{BdG} = H_t + \sum_{\vec{q}} \Psi^\dagger (\vec{q})( M + M' )  \Psi(\vec{q})
 \end{equation}
 where the two matrices appearing in the last term are
 \begin{equation}
 M' = i (Re[\Delta ' ] \tau_1 + Im [ \Delta ' ] \tau_2 ) \otimes \sigma_0 \otimes i \gamma_0 \gamma_3 ) H_D,
 \end{equation}
 and
 \begin{widetext}
 \begin{equation}
  M= [  (R_+ \tau_2 + I_+ \tau_1) \otimes \sigma_2  + (I_- \tau_2 -R_- \tau_1)\otimes \sigma_1
  + ( X \tau_1 - Y \tau_2 )\otimes \sigma_3 ] \otimes \gamma_0 ,
 \end{equation}
 \end{widetext}
 where
 \begin{equation}
 R_\pm = \frac{1}{2} ( Re( \Delta_{\uparrow})  \pm Re( \Delta_{\downarrow})  ),
 \end{equation}
 \begin{equation}
 I_\pm = \frac{1}{2} ( Im( \Delta_{\uparrow})  \pm Im ( \Delta_{\downarrow})  ),
 \end{equation}
 and
 \begin{equation}
 \Delta = X+ iY.
 \end{equation}

Before proceeding with the diagonalization of the BdG Hamiltonian it is worth pausing to register its symmetries. The Dirac Hamiltonian commutes with $N= \tau_3\otimes \sigma_0 \otimes I$ and $P = \tau_3\otimes \sigma_0 \otimes i\gamma_3 \gamma_5$, which in our representation stand for the particle-number operator and the  generator of translations.  It also commutes with $ I_K = \tau_0 \otimes \sigma_0 \otimes i \gamma_1 \gamma_5$, and $I_{uv} = \tau_0 \otimes \sigma_0 \otimes \gamma_2$,  when accompanied with the axis inversions $q_1 \rightarrow - q_1$, and
$q_2 \rightarrow - q_2$, respectively. The latter two operations represent the exchanges of the two
Dirac-points and the two sublattices, respectively \cite{herb-jur-roy}. $H_D$ also commutes with all three generators of rotations of electron spin, $\vec{S} = \tau_0 \otimes \vec{\sigma} \otimes I$. The matrix $M$ does not commute with $N$, $P$, and $I_{uv}$, but, for $\alpha=0$ under consideration at the moment, it does commute with $I_K$. It therefore represents a spatially {\it non-uniform} superconducting condensate, which is odd under the sublattice exchange and even under the exchange of Dirac points. We will call it the {\it s-Kekule superconductor}. Since it violates the spin-rotational symmetry, the matrix $M$ represents a triplet superconducting state, which breaks two generators of spin rotations.

The matrix $M'$, on the other hand, is a product of the Dirac Hamiltonian and another matrix which,  in our representation, by itself would represent the singlet s-wave order parameter \cite{comment}. Since the two factors anticommute the presence of the imaginary unit in Eq. (13) makes the matrix $M'$ Hermitian. The matrix $M'$ represents the hidden superconducting order \cite{uchoa}. This superconducting state, however, suffers from an energetic disadvantage: since $M'$ vanishes precisely at the Dirac points, opening of the order parameter $\Delta'$ seems like an ineffective  way to lower the energy of the filled Dirac-Fermi sea. We will argue shortly that in competition with the Kekule triplet the  hidden order is likely to be energetically inferior. Therefore we set $\Delta'=0$ for the time being, to return to the issue of the hidden superconducting order only after we determine the optimal triplet s-Kekule order parameters.

Setting then $\Delta' =0$ one finds
 \begin{equation}
 (H_D + M)^2 = (q^2 + m^2) (\tau_0 \otimes \sigma_0 \otimes I) + 2 \tau_3 \otimes (\vec{n} \cdot \vec{\sigma}) \otimes I,
 \end{equation}
 where the vector $\vec{n}$ has the components:
 \begin{equation}
 \vec{n} = (X R_+ + Y I_+, Y R_- - X I_-, R_+ R_- + I_+ I_-),
 \end{equation}
 and the mass-gap is
 \begin{equation}
 m^2 = X^2 + Y^2 + R_+ ^2 + I_+ ^2+ R_- ^2 + I_- ^2.
 \end{equation}
 The mean-field ground state energy per site of honeycomb lattice is therefore
 \begin{equation}
 \frac{E}{2N} = \frac{3 m^2}{ 2 V} - \sum_{ s=\pm } \int \frac{ d\vec{q} }{ ( 2\pi)^2 } ( q^2 + m^2 + 2 s |\vec{n}|) ^{1/2},
 \end{equation}
 where $N$ is the number of points in the first Brillouin zone. We also assume an ultraviolet cutoff $\Lambda$ in the integral over momenta, which is here performed only near the two Dirac points. Differentiating with respect to $|\vec{n}|$ immediately shows that for any value of the mass $m$ the minimum of energy lies at $|\vec{n}|=0$. We set therefore $n_1= 0$ and $n_2 =0$. Viewed as a set of two linear equations for the  variables $X$ and $Y$ they will have a non-trivial solution only if
 \begin{equation}
 R_+ R_- + I_+ I_- =0,
 \end{equation}
 which also happens to be the remaining equation $n_3 =0$. The trivial solution $X=Y=0$ will be discussed separately in sec. VII. The condition  $|\vec{n}|=0$ yields therefore only two, and not three independent equations. The last equation then implies
 \begin{equation}
 |\Delta_ \uparrow|= |\Delta_ \downarrow|
 \end{equation}
 and the remaining condition constrains the order parameter's phases as
 \begin{equation}
 \phi_\uparrow + \phi_\downarrow = 2 \phi + \pi,
 \end{equation}
 where $\Delta_{\sigma } = |\Delta_{\sigma} | \exp ( i \phi_{\sigma}) $, and $\Delta = |\Delta| \exp ( i \phi) $. Finally, the minimum of energy is at the value of the mass-gap $m=m_0 $ determined by the gap equation:
 \begin{equation}
 1= \frac{2 V}{3} \int \frac{ d\vec{q} }{ ( 2\pi)^2 }  \frac{1}{( q^2 + m_0 ^2) ^{1/2}  } .
 \end{equation}
 which has a  solution for $V>V_c$. At the minimum, after some straightforward algebra the matrix $M$  can be written as:
 \begin{widetext}
 \begin{equation}
 M = m_0 (\tau_1 \cos \phi  - \tau_2 \sin \phi ) \otimes [ \sin \theta ( \sigma_1 \cos (\phi_\downarrow - \phi)  + \sigma_2 \sin  (\phi_\downarrow - \phi) ) +  \sigma_3 \cos \theta  ] \otimes \gamma_0,
 \end{equation}
 \end{widetext}
 where $|\Delta| = m_0 \cos\theta$, and  $|\Delta_\uparrow| = |\Delta_\downarrow|= m_0 \sin\theta$.
 At the minimum of the energy the Kekule state has three hard and three soft modes:
 the angles $(\theta, \phi_\downarrow -\phi)$ determine the preferred spin axis for the triplet state, and $\phi$ is the superconducting phase. Note that the condition  for the energy minimum $|\vec{n}|=0$ eliminated three out of six linearly independent matrices that appear in the matrix $M$ in Eq. (14). The remaining three matrices anticommute among themselves as well as with the Dirac Hamiltonian and therefore enter as a sum of squares into the expression of the ground state energy. This quite generally appears to be the optimal way for the filled Dirac-Fermi sea to lower its energy. Another example of this rule is the emergence of the easy plane for the N\'{e}el order parameter for the antiferromagnetic  state on the honeycomb lattice in the magnetic field \cite{so3}. Further consequences of this rule for the form of the order parameter in the presence of the terms that break rotational symmetry will be discussed in sec. VII.

 \section{Hidden order parameter}

   Let us now restore the possibility of the hidden superconducting order, while retaining the energy-minimum condition $\vec{n}=0$. The mean-field energy per site is now modified into:
   \begin{equation}
 \frac{E}{2N} = \frac{3 (m^2 + 2 |\Delta'|^2)  }{2V} -  2 \int \frac{ d\vec{q} }{ ( 2\pi)^2 } [ q^2 (1+ |\Delta'|^2) + m^2 ] ^{1/2}.
 \end{equation}
 In writing this expression we assumed the relative phase between the hidden and the Kekule order parameters to be  $\pi/2$, so that the matrices $M'$ in Eq. (13) and $M$ in Eq. (26) anticommute and enter the energy expression as a sum of squares. The relative factor of two in the first term derives from the sum of order parameters over a Kekule unit cell (Fig. 1). The critical interaction for the appearance of the s-Kekule order is therefore $V_{c}= 3  \pi/\Lambda$, whereas for the hidden order, in absence of the Kekule state, {\it it would be} $ V' _{c}= 18  \pi/\Lambda^3$. Choosing the cutoff $\Lambda$ even as big as unity, which, for instance, would represent  the interval of the energies over which the tight-binding density of states is approximately linear, we see that  by increasing the interaction at $V=V_c$ the system first becomes the Kekule superconductor, with $m_0 \neq 0$. Upon further increase of the interaction the amplitude of the order parameter $m_0$ grows, and then suppresses any  appearance of the hidden order. We believe the reason for this outcome of the competition to be quite physical: given the choice whether to open the gap in spectrum or increase the velocity of excitations, all the rest being equal, the system chooses the former option as energetically preferable. The reader should be warned, however, that this conclusion could in principle be overturned upon inclusion of the states farther from the Fermi level into the energy calculation. The pure hidden order, or even the coexistence of the two orders, seem conceivable as well. Since the presence of the residual repulsive interactions in a real system will always broaden the single-particle states away from the Fermi level, it is difficult to say anything more definite on this issue beyond the low-energy approximation we employed.

We have checked, nevertheless, that our conclusion remains unaltered within the present mean-field calculation that keeps {\it all} quasiparticle states perfectly sharp, upon inclusion of the states from the entire first Brillouin  zone. This way we find the two critical interactions defined above to be $V_c ' = 3/0.786= 3.816$ and $ V_c = ((3/2)/ 0.727)= 2.063$,  in qualitative agreement with the conclusion based on the linear approximation to quasiparticle dispersion. For further details of this computation the reader should consult the Appendix.

\section{p-Kekule state}

We turn to a general Kekule state with the parameter $\alpha\neq 0$ next. Select the spin axis so that $\Delta_{\uparrow} = \Delta_{\downarrow} =0$ by setting the angle $\theta =0$ in Eq. (26).  Without a loss in generality one may choose then the order parameter $\Delta$ to be real, and write the BdG Hamiltonian {\it in real space} as
\begin{equation}
    H_{BdG} = \sum_{ \vec{x}_1, \vec{x}_2} \Phi^\dagger (\vec{x}_1) [ (\tau_0 \otimes T) + \Delta (\tau_1 \otimes K) ] \Phi(\vec{x}_2),
\end{equation}
where $ \vec{x}_1 $ and  $ \vec{x}_2 $ belong to the same sublattice, and
\begin{equation}
\Phi^\top (\vec{x}) = (u_\uparrow (\vec{x}), v_\uparrow (\vec{x}+\vec{b}) , u^\dagger _\downarrow (\vec{x}), - v^\dagger _\downarrow (\vec{x}+\vec{b})).
\end{equation}
$\vec{b}$ is one of the three  vectors that connect the nearest neighbors of the honeycomb lattice. The elements of the connectivity matrices $T$ and $K$ represent the uniform and Kekule hopping integrals between the nearest-neighbors, respectively. By rotating $\tau_1$ in the second term into $\tau_3$ then, we find that the energy of the Dirac-Fermi sea in presence of a superconducting Kekule order parameter $K$ is given by the sum of the energies of the {\it two}  copies of the Dirac-Fermi seas for the spinless fermions: one  in presence of the Kekule hopping pattern $+K$, and the other in the pattern $-K$. We have therefore computed the energy $f(\alpha)$ for the single copy as a function of the parameter $\alpha$ at various values of the amplitude $|\Delta|$. The typical result is depicted in Fig. 2. The function $f(\alpha)$ may be shown in general to be even, and periodic with the period $2 \pi/3$, which reflects the rotational symmetry of the honeycomb lattice. The computation shows that its absolute minimum is always at $\alpha=0$, in agreement with the recent work \cite{weeks}, and the maximum at $f(\pi/3)=f(\pi)$. We then find that $2 f(\pi/2) < f(0) + f(\pi)$, as long as $|\Delta|< 2.725$. The transition from the semimetallic phase is therefore into the superconducting Kekule phase with $\alpha=\pi/2$, which we therefore name {\it p-Kekule}. For $|\Delta|>2.725$, deep within the superconducting phase, we find $\alpha=0$ solution to eventually become energetically favorable, with a discontinuous transition between the s-Kekule and p-Kekule superconductors.

\begin{figure}[t]
{\centering\resizebox*{80mm}{!}{\includegraphics{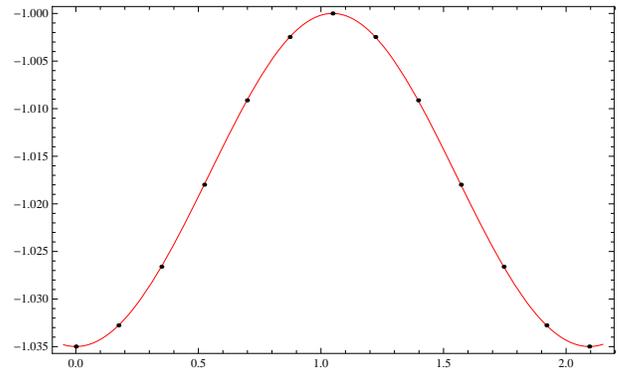}}
\par} \caption[] {The energy per site $f(\alpha)$ of the Dirac-Fermi sea of spinless fermions, hopping between nearest neighbors of the honeycomb lattice with the Kekule hopping amplitude $ 1+|\Delta|\cos ( \vec{Q}\cdot (\vec{x}+\vec{y}) +\alpha) $ for $ |\Delta| = 1$, as a function of the parameter $\alpha$. The precise type of the Kekule superconducting order depends on the sign of the combination $f(0) + f(\pi/3) - 2 f(\pi/6)$. The points  are the computed values, and the red line is our best fit $-1.018 - 0.0175 \cos(3\alpha)  + 0.000248 \cos(6 \alpha) +O(10^{-5} \cos(9 \alpha))$. This implies the p-Kekule order (see the text). The transition into the s-Kekule state at $|\Delta|=2.725$ essentially  corresponds to the change in sign of the second harmonic of this function. }
\end{figure}

\section{Other superconducting states}

 Let us also recognize the other gapped superconducting states, as the possible mass-terms that anticommute {\it both}  with the Dirac Hamiltonian $H_D$ and with the number operator $N$:

 a) the standard s-wave superconductor with the on-site pairing,
\begin{equation}
    \langle \Psi^\dagger  [( \tau_1 \cos \phi  + \tau_2 \sin \phi )  \otimes \sigma_0 \otimes i\gamma_0 \gamma_3 ]  \Psi \rangle,
    \end{equation}
which is translationally   invariant, even under the valley and/or sublattice exchange, but odd under the exchange of spin labels (spin singlet).

b) the f-wave   \cite{honerkamp}
\begin{equation}
    \langle \Psi^\dagger  [  ( \tau_1 \cos \phi  + \tau_2 \sin \phi )   \otimes \vec{\sigma} \otimes i\gamma_0 \gamma_5 ] \Psi \rangle,
     \end{equation}
     which is translationally invariant, even under the sublattice and spin exchanges (spin triplet), but odd under valley exchange.

c) the p-Kekule state discussed in the previous section written explicitly is
     \begin{equation}
  \langle \Psi^\dagger  [( \tau_1 \cos \phi  + \tau_2 \sin \phi )  \otimes \vec{\sigma} \otimes i\gamma_1 \gamma_2 ] \Psi \rangle.
     \end{equation}
One can further  construct all the gapless (hidden) condensates, as
\begin{equation}
i \langle \Psi^\dagger  M H_D \Psi \rangle,
\end{equation}
where $M$ is a mass-matrix for {\it any} of the above gapped superconducting states. Choosing the matrix $M$ to correspond to the s-wave superconductor yields the original hidden order of ref. (4). Since the Hermitian matrix $i M H_D$ by construction then anticommutes with the Dirac Hamiltonian $H_D$ while at the same time being proportional to it, its addition to $H_D$ will effectively only renormalize the velocity of the Bogoliubov excitations, as manifest in Eq. (27).

  Of course, one can imagine many other matrices that do not commute with the number operator, and will therefore represent some superconducting order, which nevertheless do not fall into any of the categories listed above. These fail to anticommute with the Dirac Hamiltonian, and as such neither gap out, nor increase the velocity of the Dirac fermions. Development of these order parameters would not therefore be particularly energetically advantageous, which is the reason behind their omission here.

\begin{figure}[t]
{\centering\resizebox*{70mm}{!}{\includegraphics{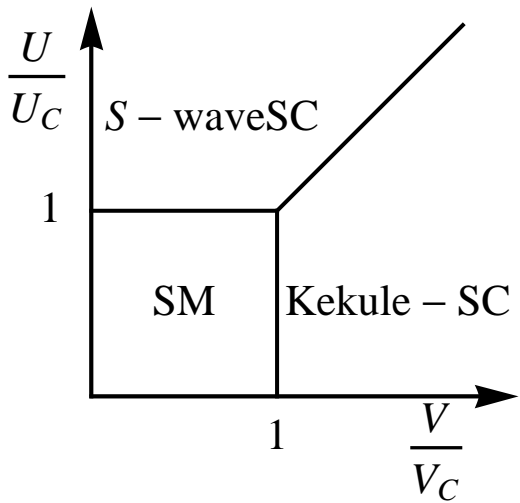}}
\par} \caption[] {The schematic $T=0$ phase diagram in the model with the on-site ($U$) and the nearest-neighbor ($V$) attractions. At the boundary between the two superconducting states the order parameter acquires the larger $O(5)$ symmetry, as the energy becomes invariant under the rotations of the Kekule into the s-wave order with the relative phase of $\pi/2$. }
\end{figure}

 One can expect the gapped superconducting states to compete in the phase diagram for attractive interactions in a close parallel with the competition between insulators when the interactions are repulsive \cite{herb-jur-roy}. As an illustration, in Fig. 3 we present the mean-field phase diagram in presence of both the on-site and the nearest-neighbor attractions at half filling. In analogy with the insulating case, there is a discontinuous transition between the ordered phases, whereas the transitions out of  the semimetallic phase may be expected to be  continuous \cite{metzner, herb-jur-vaf}. One novel feature is that because of the $U(1)$ symmetry in the superconducting phase the matrices representing different superconducting states can be chosen so as to {\it anticommute}. Consider the above p-Kekule state with the phase $\phi=0$ and the spin axis along z-direction, for example. Choosing the uniform s-wave state with $\phi=\pi/2$ makes the two representative mass-matrices anticommuting, so that right at the boundary between the two phases the system acquires a larger symmetry $O(5)$. Adding the third axis for the  second-nearest neighbor attractive interaction introduces then a region of the f-wave order, with the discontinuous transitions between any two of  the three phases. Interestingly, there is a unique anticommuting f-wave state,  with $\phi=\pi/2$  and with the spin axis along z-direction, that may be added to the above combination of the already anticommuting Kekule and s-wave order parameters. At the point in the phase diagram where the three phases would meet, an even larger, $O(6)$,  symmetry emerges.

 If the preferred non-uniform superconducting state is the s-Kekule, on the other hand, the possible anticommuting  states are again the uniform s-wave and f-wave condensates, but this time both with the same phase as the one of the s-Kekule state.

\section{Topology and defects}

In the ordered phase, the mass-matrix for the Kekule order parameter in Eq. (26) lives on the $S_1\times S_2$ target space, but with opposite points identified. In other words, the order parameter space is the product of $S_2$ for the spin direction and half of $S_1$ for the superconducting phase, which is equivalent to $S_3$, the sphere in four dimensions. That this is indeed the target space becomes clear upon recalling that the minimum with $\vec{n}=0$, the mean-field free energy in Eq. (21) depends only on the mass  $m^2 = |\Delta|^2  + |\Delta_\uparrow|^2 $, where the two complex order parameters $\Delta$ and $\Delta_\uparrow$ are constrained only by the condition that $m$ is  the solution of the gap equation.

Since the first and the second homotopy groups of $S_3$ are trivial, $\pi_1 (S_3) = \pi _2 (S_3) = 1$, there are no stable topological defects, and the massless fluctuations in the ordered phase should be correctly described by the $O(4)$ non-linear sigma model \cite{book}. We therefore do not expect a true finite temperature phase transition from a semimetal into the Kekule superconductor, but only a crossover when the superconducting correlation length $\xi \propto \exp(c m_0/T)$, with $c$ as a (non-universal) numerical constant, reaches the size of the sample \cite{oshikawa}.

 A reduction of the rotational symmetry would change the target space and allow stable vortex excitations. Let us consider the case of a possible easy plane first. Such an anisotropy  may be introduced most simply by placing the Kekule superconductor into a magnetic field. The Zeeman term representing the coupling of the magnetic field to the electron spin  will be proportional to the generator of rotations along the direction of the magnetic field, $\tau_0 \otimes \sigma_3 \otimes I$, for example. Since this matrix commutes with the Dirac Hamiltonian and with the third term in the Kekule mass-matrix $M$ in Eq. (14) that is proportional to $\Delta$, while it anticommutes with the two other terms in $M$ that are proportional to $\Delta_\sigma$, the minimization of the energy in the presence of Zeeman coupling is formally equivalent to the problem of  N\'{e}el ordering in graphene catalyzed by the magnetic field \cite{so3, herbutAF}. The result is that $\Delta =0$, since that way the Kekule mass matrix anticommutes with the Zeeman term. The way to understand this physically is to realize that in such a state the spins of paired electrons are all orthogonal to the magnetic field, so it becomes easier for them to tilt and provide a finite magnetization in the field direction. The minimum condition $\vec{n}=0$ then translates into $|\Delta_\uparrow |= |\Delta_\downarrow|$, the same as without the magnetic field, but without a further constraint on the phases of the two complex order parameters. The s-Kekule mass-matrix in Eq. (14) for such an easy plane may be then rewritten differently as
\begin{widetext}
 \begin{equation}
 M = m_0 (\tau_2 \cos \frac{\phi_\uparrow+\phi_\downarrow}{2}  + \tau_1 \sin \frac{\phi_\uparrow + \phi_\downarrow}{2}) \otimes ( \sigma_2 \cos \frac{\phi_\uparrow - \phi_\downarrow}{2}  + \sigma_1 \sin \frac{\phi_\uparrow - \phi_\downarrow}{2} ) \otimes \gamma_0.
 \end{equation}

 \end{widetext}
 The target space with the easy plane anisotropy is thus $S_1 \times S_1 $, with the factors corresponding to the two phases $\phi_\uparrow$ and $\phi_\downarrow$. Since the first homotopy group of $S_1$ is non-trivial, $\pi_1 (S_1) = Z$, there are different types of topologically distinct vortex excitations. For example, winding just one of the phases by $2\pi$ causes both $(\phi_\uparrow+\phi_\downarrow)/2$ and $(\phi_\uparrow-\phi_\downarrow)/2$ to change from zero to $\pi$, i. e. both the first, phase term, and the second, spin-axis term in the above matrix make half a circle. This is sometimes referred to as ``half-vortex" \cite{woelfle}. On the other hand, winding both the phases $\phi_\uparrow$ and $\phi_\downarrow$ in the same sense by $2\pi$ leaves the angle of the spin-axis intact, and produces the standard full vortex in the superconducting phase. Finally, winding the two phases  $\phi_\uparrow$ and $\phi_\downarrow$ in the opposite sense by $2\pi$ produces a third type of vortex, this time in the direction of the spin-axis only.

 An easy axis, on the other hand, is introduced by the spin-orbit coupling, for example. Consider adding a weak perturbation to the Dirac Hamiltonian proportional to $\tau_3 \otimes \sigma_3 \otimes i\gamma_1\gamma_2$, which in  our representation corresponds to the third component of the spin-triplet version of the time-reversal symmetry breaking mass, introduced by Haldane \cite{haldane} and discussed in the context of spin-orbit interaction in graphene by Kane and Mele \cite{kane}. The presence of such a term would again select the piece of the Kekule mass matrix that anticommutes with it, but due to the $\tau_3$ matrix in the first, Nambu's factor, this now implies that $\Delta_\uparrow= \Delta_\downarrow=0$. The s-Kekule matrix assumes the form as in Eq. (26), with $\theta=0$. The target space therefore in this case reduces to the usual $ S_1 $, with only the standard vortices as the topological excitations. The internal structure of such a vortex has been studied in ref. \cite{herbut}.

 Finally, it should be understood that we discussed the breaking of the rotational symmetry in the s-Kekule state for simplicity only, and that everything said applies equally to the p-Kekule state as well.

\section{Summary and discussion}

  To summarize, we  introduced the non-uniform superconducting state on graphene's honeycomb lattice, and argued that it is the mean-field solution of the simple model with nearest-neighbor attraction. The order parameter for this state lives on the bonds of the lattice, and forms the Kekule lattice with the period three. Competition between different such Kekule superconducting states, as well as between the Kekule and the other possible gapped and gapless superconducting states was discussed. The Kekule superconductor has a spin-triplet order parameter, which lives on the surface of the $S_3$ sphere. Target spaces for the order parameter and the topological defects in presence of some simple symmetry breaking terms were determined.

A Kekule insulator which breaks the translational invariance of the honeycomb lattice has been previously proposed and discussed in literature \cite{chamon, hou}. However, it appears that this state is not the ground state of the simplest  model with only the nearest-neighbor repulsion, since there is an energetically superior charge-density-wave that breaks the sublattice exchange symmetry available. This should be contrasted with the situation for the attractive interactions considered here, where the competing superconducting state is the gapless superconductor, which we argued should have a higher energy. It was argued recently, however, that the Kekule insulator does become the mean-field ground state when there is a balance between the nearest-neighbor and the second-nearest-neighbor components of the repulsive interactions \cite{weeks}.

We described here only the problem at half-filling in detail, where a finite interaction is needed to cause the superconducting transition.  At a finite chemical potential  the density of states at the Fermi level also becomes finite, and there is the usual BCS instability at an infinitesimal attraction. For small deviations from the half-filling, however, the symmetry of the superconducting state is essentially determined by the solution at the Dirac point. For the nearest-neighbor attraction as the dominant component of the interaction one should therefore expect the non-uniform Kekule state we discussed to persist at a finite doping  as well. As long as the Fermi surface around the Dirac points stays circular \cite{brief} the states with the momenta $\vec{Q}+ \vec{q}$ and $\vec{Q}-\vec{q}$ may both be near the Fermi surface and be paired up by the Kekule order parameter with the momentum $2\vec{Q}$, essentially the same way as right at half-filling.

\section{Acknowledgement}

This work was supported by the NSERC of Canada. We thank C.-K. Lu and M. Franz for useful discussions. I. F. H. is grateful to the Institute for Solid State Physics of the University of Tokyo where a part of this work was performed for their hospitality.

\section{Appendix}

Here we determine the the susceptibilities for the hidden and Kekule orders used at the end of sec. IV, evaluated over the whole Brillouin zone. For the hidden order the energy per site may be written as
\begin{equation}
\frac{E(\Delta ') }{2N}= (\frac{3}{V}- \frac{1}{2 N} \sum_{\vec{k}} |f(\vec{k} )|)  |\Delta '|^2 + O( |\Delta '|^4)
\end{equation}
where
\begin{equation}
f(\vec{k} )= \sum_{i=1,2,3} e^{ i \vec{k} \cdot \vec{b}_i }
\end{equation}
and $\vec{b}_i$ are the three vectors connecting the nearest neighbors on the honeycomb lattice \cite{herb-jur-roy}. The sum over the wavevectros is performed over the entire Brillouin zone with $N$ points. We find
\begin{equation}
\frac{1}{2 N} \sum_{\vec{k}} |f(\vec{k} )| =  0.786
\end{equation}
in agreement with \cite{weeks}. This yields the value of $V_c '$ cited in the text.

For the critical interaction for Kekule order we need the energy as a function of the Kekule mass $m$ to the leading order. Diagonalizing the six-dimensional matrix given by Weeks and Franz \cite{weeks} and summing over the reduced Brillouin zone for the Kekule lattice we find
\begin{equation}
\frac{E(m)}{2N}= (\frac{3}{2 V}-  0.727 ) m ^2 + O( |m|^3 ).
\end{equation}
Note that the the electronic susceptibilities for the hidden and Kekule orders are rather close numerically, and the Kekule state wins mainly due to the geometrical factor of two in the first term in the mean-field energy.

\end{document}